\documentstyle{mn}

\newif\ifAMStwofonts

\newcommand{\ea} {{\rm et al. }}

\input psfig.sty

\ifoldfss
  \newcommand{\rmn}[1] {{\rm #1}}

  \ifCUPmtlplainloaded \else
    \NewTextAlphabet{textbfit} {cmbxti10} {}
    \NewTextAlphabet{textbfss} {cmssbx10} {}
    \NewMathAlphabet{mathbfit} {cmbxti10} {} 
    \NewMathAlphabet{mathbfss} {cmssbx10} {} 
  \fi
  \ifAMStwofonts
    \ifCUPmtlplainloaded \else
      \NewSymbolFont{upmath} {eurm10}
      \NewSymbolFont{AMSa} {msam10}
      \NewMathSymbol{\upi}     {0}{upmath}{19}
      \NewMathSymbol{\umu}     {0}{upmath}{16}
      \NewMathSymbol{\upartial}{0}{upmath}{40}
      \NewMathSymbol{\leqslant}{3}{AMSa}{36}
      \NewMathSymbol{\geqslant}{3}{AMSa}{3E}

    \fi
  \fi
\fi 

\ifnfssone
  \newmathalphabet{\mathit}
  \addtoversion{normal}{\mathit}{cmr}{m}{it}
  \addtoversion{bold}{\mathit}{cmr}{bx}{it}
  \newcommand{\rmn}[1] {\mathrm{#1}}

  \newmathalphabet{\mathbfit} 
  \addtoversion{normal}{\mathbfit}{cmr}{bx}{it}
  \addtoversion{bold}{\mathbfit}{cmr}{bx}{it}
  \newmathalphabet{\mathbfss} 
  \addtoversion{normal}{\mathbfss}{cmss}{bx}{n}
  \addtoversion{bold}{\mathbfss}{cmss}{bx}{n}
  \ifAMStwofonts
    \ifCUPmtlplainloaded \else
      \UseAMStwoboldmath
      \makeatletter
      \new@mathgroup\upmath@group
      \define@mathgroup\mv@normal\upmath@group{eur}{m}{n}
      \define@mathgroup\mv@bold\upmath@group{eur}{b}{n}
      \edef\UPM{\hexnumber\upmath@group}
      \new@mathgroup\amsa@group
      \define@mathgroup\mv@normal\amsa@group{msa}{m}{n}
      \define@mathgroup\mv@bold\amsa@group{msa}{m}{n}
      \edef\AMSa{\hexnumber\amsa@group}
      \makeatother
      \mathchardef\upi="0\UPM19
      \mathchardef\umu="0\UPM16
      \mathchardef\upartial="0\UPM40
      \mathchardef\leqslant="3\AMSa36
      \mathchardef\geqslant="3\AMSa3E
    \fi
  \fi
\fi 

\ifnfsstwo
  \newcommand{\rmn}[1] {\mathrm{#1}}

  \DeclareMathAlphabet{\mathbfit}{OT1}{cmr}{bx}{it}
  \SetMathAlphabet\mathbfit{bold}{OT1}{cmr}{bx}{it}
  \DeclareMathAlphabet{\mathbfss}{OT1}{cmss}{bx}{n}
  \SetMathAlphabet\mathbfss{bold}{OT1}{cmss}{bx}{n}
  \ifAMStwofonts
    \ifCUPmtlplainloaded \else
      \DeclareSymbolFont{UPM}{U}{eur}{m}{n}
      \SetSymbolFont{UPM}{bold}{U}{eur}{b}{n}
      \DeclareSymbolFont{AMSa}{U}{msa}{m}{n}
      \DeclareMathSymbol{\upi}{0}{UPM}{"19}
      \DeclareMathSymbol{\umu}{0}{UPM}{"16}
      \DeclareMathSymbol{\upartial}{0}{UPM}{"40}
      \DeclareMathSymbol{\leqslant}{3}{AMSa}{"36}
      \DeclareMathSymbol{\geqslant}{3}{AMSa}{"3E}
    \fi
  \fi
\fi 

\ifCUPmtlplainloaded \else
  \ifAMStwofonts \else 
    \def\upi{\pi}
    \def\umu{\mu}
    \def\upartial{\partial}
  \fi
\fi

\begin{document}

\title{3D eclipse mapping in AM Herculis systems - `Genetically
modified fireflies' }
\author[Pasi Hakala ,Mark Cropper \& Gavin Ramsay]
       {Pasi Hakala$^{1}$, Mark Cropper$^{2}$ \& Gavin Ramsay$^{2}$ \\
        $^{1}$Tuorla Observatory, V\"ais\"al\"antie 20, 21500 Piikki\"o,
        Finland, \\
        $^{2}$Mullard Space Science Laboratory, University College
        London, Holmbury St. Mary, Dorking, Surrey RH5 6NT, UK}
\date{ }

\pubyear{2002}

\maketitle

\label{firstpage}

\begin{abstract}

In order to map the 3-dimensional location and shape of the emission
originating within the accretion stream in AM Her systems, we have investigated
the possibilities of relaxing the hitherto-applied constraint of a
predetermined stream trajectory in modelling the eclipse profiles. We use
emission points which can be located anywhere in the Roche lobe of the primary,
together with a regularisation term which favours any curved stream structure,
connected at the secondary and white dwarf primary. Our results show that,
given suitable regularisation constraints, such inversion is feasible. We
investigate the effect of removing the regularisation term, and also the
sensitivity of the fit to input parameters such as inclination.

\end{abstract}


\begin{figure}
\centerline{\psfig{figure=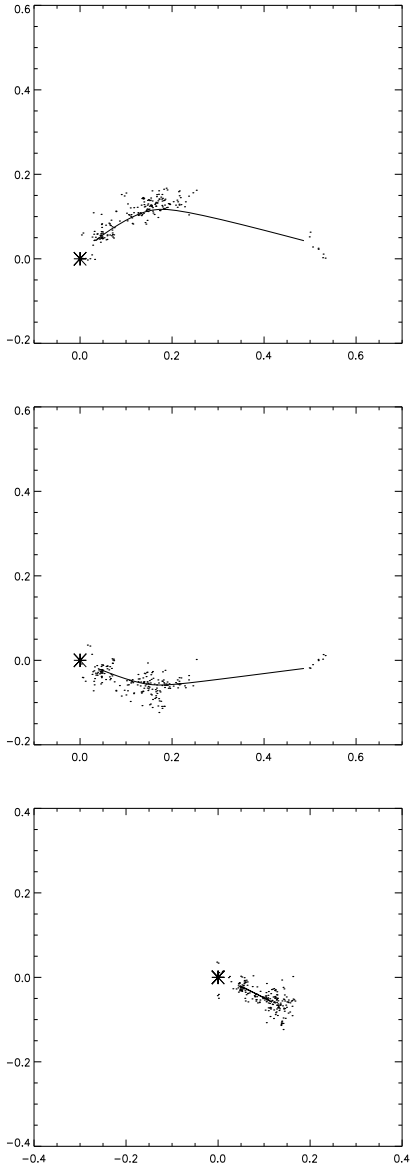,height=16.0cm,width=7.0cm,angle=0}}
\caption{A view of the swarm (Stream 3, see Figure 2.) during the
fitting process. We have overplotted the ``default path'' found for
the swarm at that moment using the SOM algorithm. The projections
(from top to bottom) are X-Y plane (orbital plane), X-Z plane and Y-Z plane.}
\end{figure}

\section{Introduction}

AM Herculis systems (or polars) are a subclass of cataclysmic
variables (CV), where a magnetic white dwarf accretes matter from
a late type dwarf secondary via Roche lobe overflow. The strong
(10-100 MG) magnetic field of the white dwarf prevents the formation
of an accretion disc, and thus the accretion proceeds through an accretion
stream, which directly impacts onto the white dwarf surface near the magnetic
poles (see Cropper 1990 for a review on polars).

Until the early 1990's it was widely accepted that almost all the
optical emission from AM Her systems originates in the cyclotron
emission region(s) near the magnetic poles of the white
dwarf. However, the discovery of highly asymmetric optical eclipse
profiles in systems like HU Aqr 
(Hakala \ea 1993, Schwope \ea 1993) demonstrated that the accretion 
stream could
contribute as much as 50\% of the total optical emission. Since then
a number of attempts have been made to use the eclipse profile
information of the AM Her systems to reconstruct the location and
other properties of the stream emission. These earlier attempts
include those of Hakala (1995), Harrop-Allin, Hakala \& Cropper (1999) and
Kube, G\"ansicke \& Beuermann (2000). Although considerable 
improvements have been made to the original concept of fitting a 1-dimensional
brightness distribution along the accretion stream path (Vrielmann \& Schwope,
2001), all the existing techniques use a pre-determined path for the
accretion stream.
In this paper we will introduce a new inversion scheme, where this
limitation has been removed. This allows us to study the stream
emission in AM Her systems in more objective manner.

\section{The Model}

All existing techniques have assumed an accretion stream which is an arc of
varying opening angle connecting the L1 point to the white dwarf 
(Hakala 1995) or more realistic ones combining the theoretical free
fall and magnetically dominated trajectories (Harrop-Allin, 
Hakala \& Cropper 1999; Kube, G\"ansicke \& Beuermann 2000, Vrielmann
\& Schwope 2001). 
However, there is observational evidence,
for instance in HU Aqr (Hakala et al. 1993) when it was in a low
accretion state, the accretion stream was visible for a longer period
of time after the start of the eclipse compared to its high accretion
state: this extended period of visibility cannot be reproduced by such
models.
This suggests that it is important to eliminate the constraints imposed by
adopting a pre-fixed trajectory. 

In our new model, the accretion stream is constructed from 200 individual
emission points which are free to move within the Roche lobe of the
white dwarf primary. Each emission point (or `firefly') will have an
angle-dependent brightness as follows: 

\begin{equation}
F_{\rmn fly}(\alpha) = F_{0}+A {\rmn cos}(\alpha),
\end{equation}
  
where $F_{\rmn fly}(\alpha)$ is the flux emitted by a single firefly at angle
$\alpha$ between the observer and the white dwarf as seen from the 
firefly, $F_{0}$ is the minimum brightness of a firefly ($F_{0} > A$, so
that $F_{\rmn fly}(\alpha) > 0$ for any value of $\alpha$) and $A$ is the 
amplitude of $\alpha$ dependence. All of the flies have the same $F_{0}$. 
In most of our modelling we have 
used $F_{0} = 3.0$ and $A = 1.0$ {\it i.e.} the `back' side of the firefly is
half the brightness of the `front' side of it. This viewing angle
dependence was introduced in order to mimic the effects of X-ray
heating and/or optical thickness of the stream, which will
produce such a phase-dependent effect on the stream emission properties. 
We will discuss the effects of this choice later in this paper.

Now that we have the emission law for a single firefly, we can
compute the integrated emission expected from any
`swarm' of fireflies. If we then add the cyclotron emission component
from the white dwarf surface (point source) and the Roche lobe filling 
secondary star, which acts purely as an obstructive element in the
modelling, we are in position to generate eclipse profiles. For the moment we
neglect the contribution of the white dwarf which is generally negligible at
optical wavelengths, unless the system is in a faint state.

\begin{figure*}
\centerline{\psfig{figure=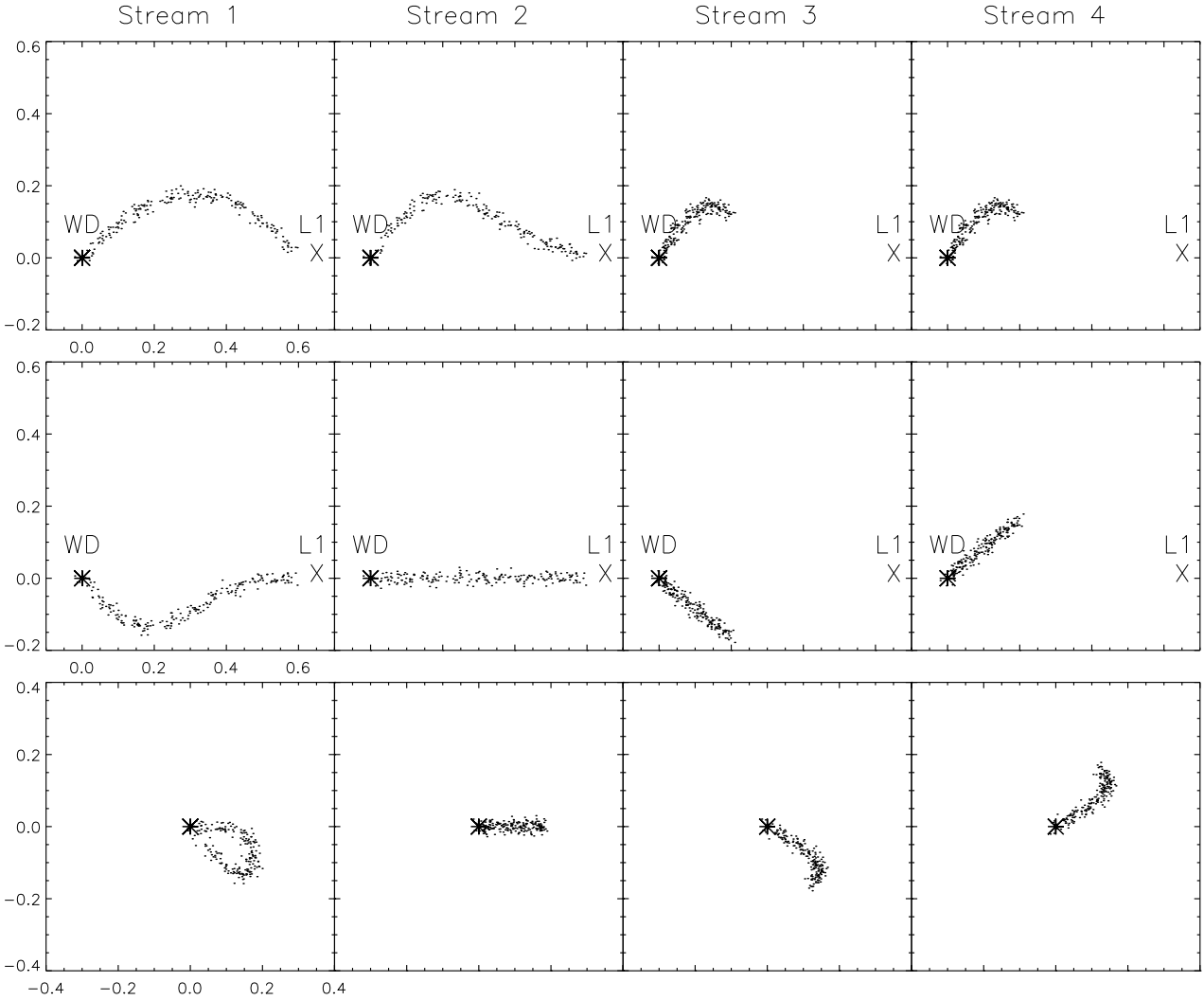,height=15.0cm,width=18.0cm,angle=90}}
\caption{The four synthetic streams used to test the method.
The different columns represent different streams, whilst the
different rows show different projections (from top to bottom:
X-Y plane, X-Z plane and Y-Z plane, where X-Y is the orbital plane
and the X axis points from the white dwarf towards the L1 point.)}
\end{figure*}

\begin{figure*}
\centerline{\psfig{figure=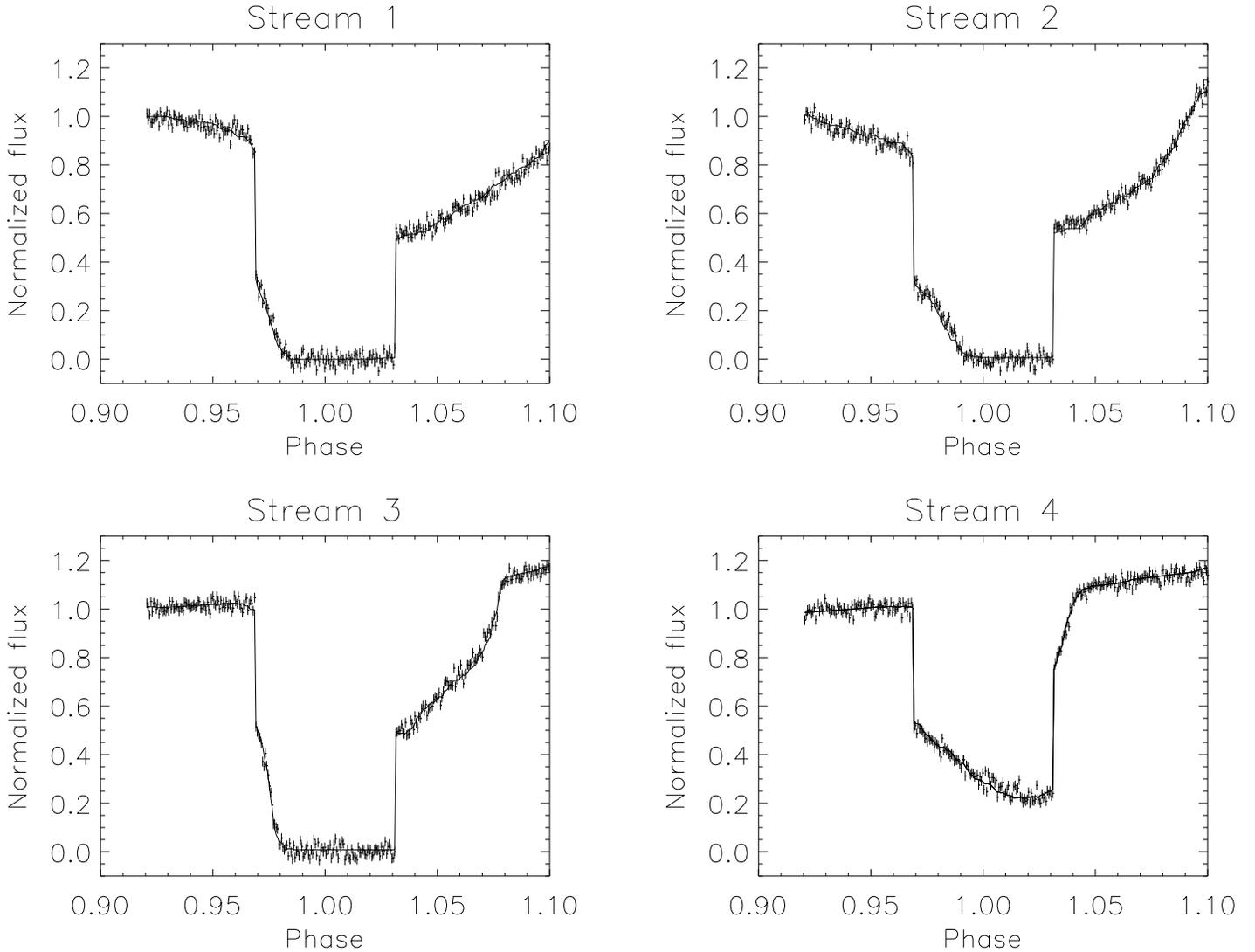,height=15.0cm,width=20.0cm,angle=90}}
\caption{The fits to the synthetic data. The four fits correspond to
the four test sets plotted in Figure 2.}
\end{figure*}

\begin{figure*}
\centerline{\psfig{figure=bestswarms.ps,height=15.0cm,width=18.0cm,angle=90}}
\caption{The best fit reconstructions of the synthetic streams shown
in Figure 2. The panels are the same as in Figure 2.}
\end{figure*}
    
\subsection{Regularisation}

Finding the best swarm of fireflies that can
fit the data will generally not produce a unique answer. For 200 fireflies we
have 600 free parameters to fit, and our eclipse
profiles may not contain that many datapoints. Even if they do, the structure
of the swarm may be too complex to recreate from the 'two cuts' available at
eclipse ingress and egress. Thus
some sort of regularisation is required in order to generate
unique solutions. 

Even with a 
regularisation term there will probably not be a unique set of 200
locations in the parameter space defining the best swarm. Instead,
the 200 fireflies act as markers defining the location and shape of
the emitting region. Thus the solutions are unique in the
sense of location and shape for the resulting swarm. This reduces the effective
dimensionality of the problem from 600 (in this case)
to $\sim$dozen. One can roughly estimate the effective number of free
parameters, by considering that in a 'banana' shaped
swarm the effective free parameters would be the locations of both
ends of the banana (3+3 parameters), some measure of curvature (1-2
parameters), and the thickness of the banana at, say 3-5, locations
along it.
This would thus result in 10-13 free parameters, if we tried to
fit a banana shape for a stream instead of using a swarm.   

With this in mind, we have decided on the following regularisation criteria for
our swarms. Firstly, in accordance with the aim of allowing a free stream
trajectory, we impose as minimal a restriction on the solutions as
possible. However, we do know that the stream should start from the L1
point and it should impact on the white dwarf surface. For the regularisation
itself, we have chosen to prefer swarm shapes that have minimal distance to 
a best fitting curve that goes through them. This is physically realistic,
since in the case of both the ballistic and magnetically confined
trajectories, the stream forms a funnel like structure.

We proceed by finding, for each swarm, the best fitting
curve that will start from close to the L1 point, pass through
the swarm and end near the white dwarf. The regularisation term is then
the squared sum of the minimum distances to this curve for all the
fireflies and will a have minimum for banana shaped swarms 
(see eq. A4 in Appendix A). Such shapes approximate 
what might be physically expected. 

In order to
find such a curve for any given swarm, we have
used a type of unsupervised neural network called Self-organizing Map
(SOM, Kohonen 1990). This algorithm is a topology-conserving 
clustering algorithm, which will classify any dataset into a predefined
number of clusters. In our regularisation application, the locations
of the flies form the dataset and the centres of these clusters define 
the curved line through the swarm and will provide the reference
points to which the minimum distances are
determined. More details of Self-organizing maps, together with
further discussion, is given in Appendix A.

\subsection{Fitting}

The actual fitting procedure is carried out using a variant of genetic
algorithms (GA), which have proven successful in a number of
astronomical applications (Hakala 1995, Potter, Hakala \& Cropper
1998, Harrop-Allin, Hakala \& Cropper 1999, Charbonneau 1995
(review)). We give a brief description of our GA procedure in 
Appendix B. 

The fitting consists of minimizing a merit function 
$F$ by evolving the swarm shapes. The merit function $F$ is defined by
\begin{equation}
F = \sum_{\rmn i=1}^{\rmn n}[{{D_{\rmn i}-M_{\rmn i}}\over\sigma_{\rmn
i}}]^2 + \lambda S_{\rmn reg}
\end{equation}
where the first part is the normal $\chi^2$ part of the fit ($D_{\rmn i}$
are the data points, $M_{\rmn i}$ the corresponding model values and
$\sigma_{\rmn i}$ the errors on the data points),
$\lambda$ is the Lagrangian multiplier and $S_{\rmn reg}$ is a regularisation
term as described above (and in Appendix A). Our GA minimization is 
carried out using
typically 500 different swarms per generation of solutions, and the 
convergence occurs typically in 50-100 generations.
Note, that whilst GA is used to minimize eq. 2, the SOM algorithm is
needed for evaluation of the regularisation term ($S_{\rmn reg}$).

Figure 1. shows an interim view of a swarm during the
fitting process, as well as the current regularisation curve
for that particular swarm.
   
\section{Testing the Algorithm}

\subsection{Model fits to synthetic data}

Given the nature of the model and the fitting method, there is
probably no formal way of proving the uniqueness of the solutions. 
We therefore need to 
examine the method numerically, using synthetic datasets for
which we know the shapes and locations of the emission regions (synthetic 
swarms). Here we present four such datasets and the corresponding results 
from our inversion. 

In Figure 2 we have plotted these datasets in 
three panels, showing the X-Y, X-Z and Y-Z projections of the
datasets. The white dwarf is located at (0,0,0), and the centre of the
secondary star at (1,0,0). Points with positive Z-coordinates are
on the same side of the orbital plane as the observer for inclinations
$<90^{\circ}$. The free fall ballistic stream has a
positive Y-coordinate.

Then, given the four synthetic streams, we have computed four synthetic eclipse
profiles, to which we have added noise (2\% RMS). These profiles, with
corresponding model fits (generated using the GA and regularisation described 
above) can be found in Figure 3. The stream
structures (swarms) producing these fits are plotted in Figure 4: these can be
directly compared with the original streams in Figure 2.

We first consider the reconstructions in the X-Y plane. The synthetic
streams have been reproduced remarkably well, with only very few
points being wrongly attributed to the ballistic portion of the stream
in the case of a wholly magnetically channelled stream (streams 3 \&
4).  In the case of the X-Z and Y-Z planes, we do not expect the
reconstructions to be as good because at an inclination of 80 degrees
(as used here) the eclipse profiles are dominated by movement in the
X-Y plane. Nevertheless, the agreement is sufficient to be
encouraging and the method is capable of discriminating between lower 
pole and upper pole accretion modes.

As the regularisation term will force the ends of the stream towards
the white dwarf and the L1 point, the resulting streams are also 'rounded'
to some extent.
In this sense the regularisation acts somewhat like a rubber band,
fixed to the white dwarf and L1 point at the ends and forced to pass through 
the swarm defining the stream.  

\subsection{Stability of the solutions}

Secondly we have studied the stability of our solutions i.e. whether
the resulting swarms always have the same location and shape. This is
very easily done with our GA optimisation code, as one only needs to
start the fitting procedure with a different seed for the random
number generator. This naturally leads to different initial swarm 
populations and thus we start at different locations in the parameter
space. We have done this by fitting each of the synthetic datasets 10
times. The results can be seen in Figure 5, where
we have plotted the resulting 10 streams on top of each other. This
figure, again, is directly comparable with Figures 2 and 4.
Clearly all the solutions are comparable with each other. Thus we 
believe that our method is capable of producing unique solutions.  

\begin{figure*}
\centerline{\psfig{figure=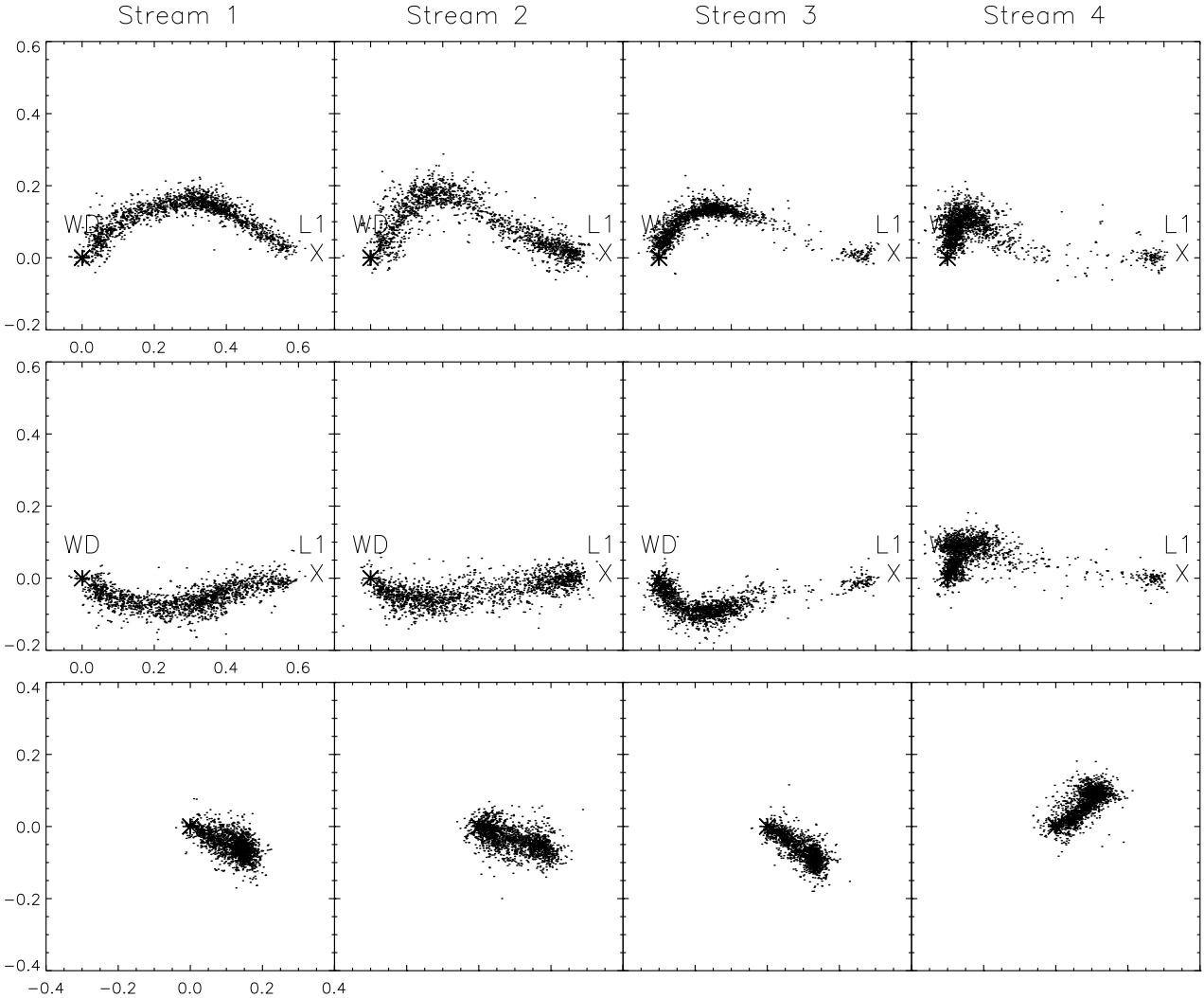,height=15.0cm,width=18.0cm,angle=90}}
\caption{The results of 10 different reconstructions of the four test
sets (see text for details). The different solutions are overplotted 
in each panel to demonstrate the stability of our solutions. The
panels are the same as in Figure 2.}

\end{figure*}

\begin{figure}
\centerline{\psfig{figure=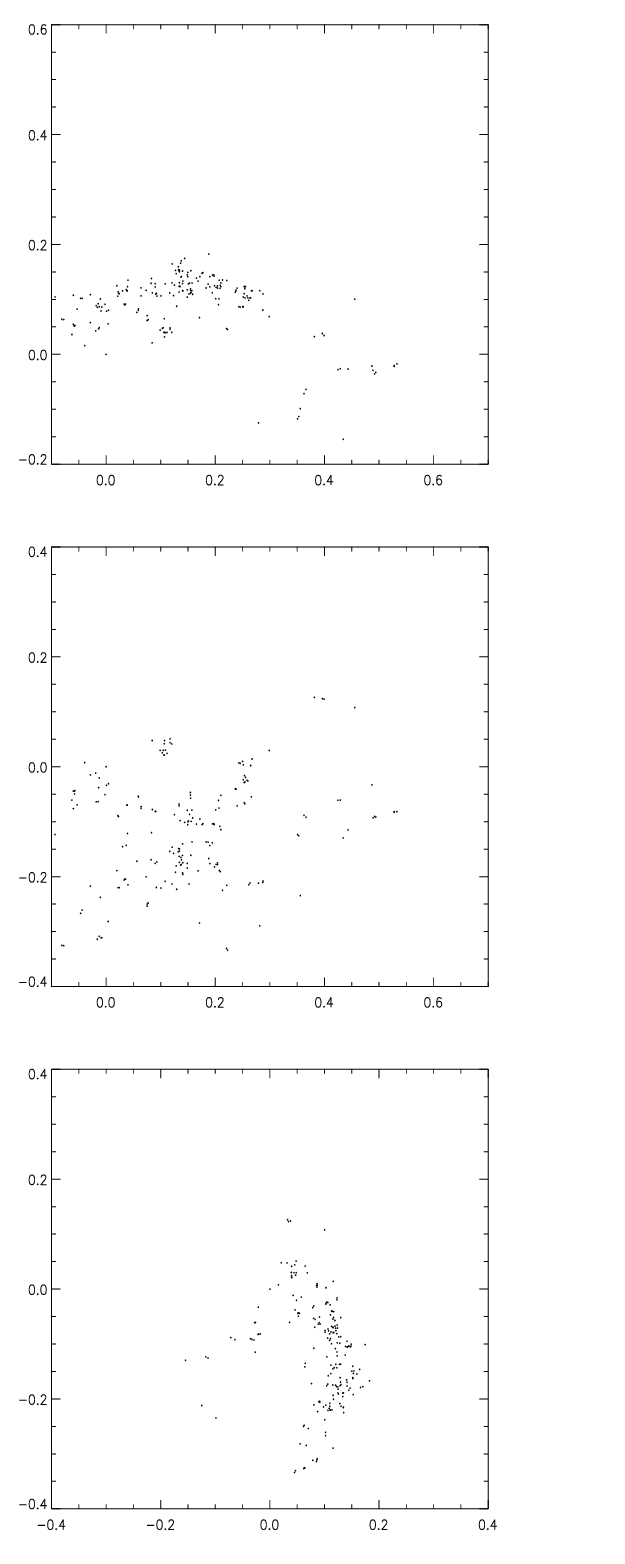,height=16.0cm,width=7.0cm,angle=0}}
\caption{The reconstruction of the third synthetic stream with no
regularisation. The panels are the same as in earlier figures.}
\end{figure}

\begin{figure}
\centerline{\psfig{figure=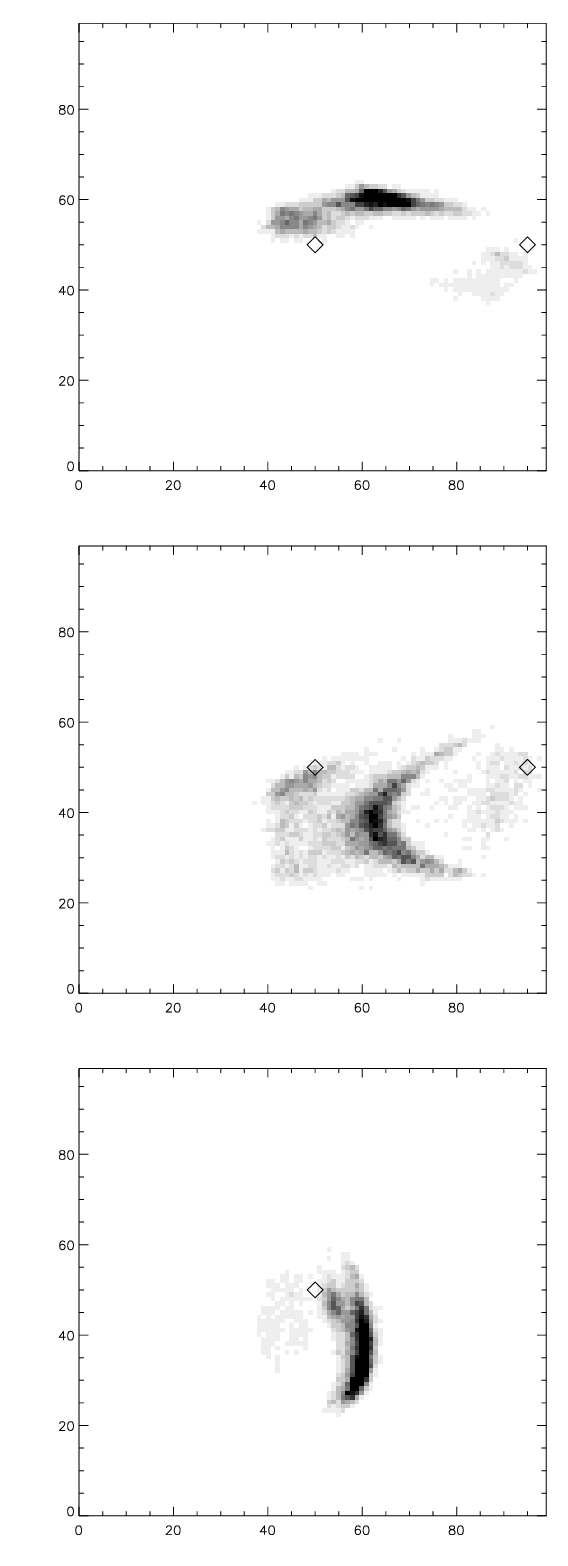,height=16.0cm,width=6.0cm,angle=0}}
\caption{200 reconstructions of the third synthetic stream with no
regularisation. This plot shows the resulting  'fly density image' in
grayscale. Note, that here the scale is larger covering most of the
primary Roche lobe. The white dwarf is located always in the centre 
(marked by a diamond). Also the L1 point is marked similarly. X and 
Y axis labelling indicate the pixel resolution used for binning of
flies. The panels are the same as earlier.}
\end{figure}

\subsection{Fits without the regularisation term}

To illustrate the effect of the regularisation we have carried
out similar fitting without any regularisation. For this purpose
we have used Stream \#3, a case where the ballistic part of the stream does 
not emit. As expected, the algorithm finds a perfectly adequate fit
to the data (not shown), 
but the stream from this fit (Figure 6) is a less faithful
reproduction of the original stream (in Figure 2). We can see that the X-Y locations are reproduced
to a fair extent, but the Z direction is less well reproduced. 

Fits without the regularisation term are nevertheless extremely
interesting. Since even when the quasi-linear structure prejudice for 
the stream is dropped, these fits represent the `minimum-assumption' 
fits to the data. Multiple fits with different seeds, as used in
Figure 7, will indicate where emission from the stream could emanate, 
while remaining consistent with the data, and where it cannot. 

We have therefore perfomed 200 fits without any regularisation for the stream
\#3. The resulting 'fly densities' are shown in Figure 7. One can
clearly see that the main effect of the regularisation term is that 
it controls the distribution of flies in Z plane. This is fairly
natural since there is less information about the Z
coordinate in the eclipse profiles. 
 
\begin{figure}
\centerline{\psfig{figure=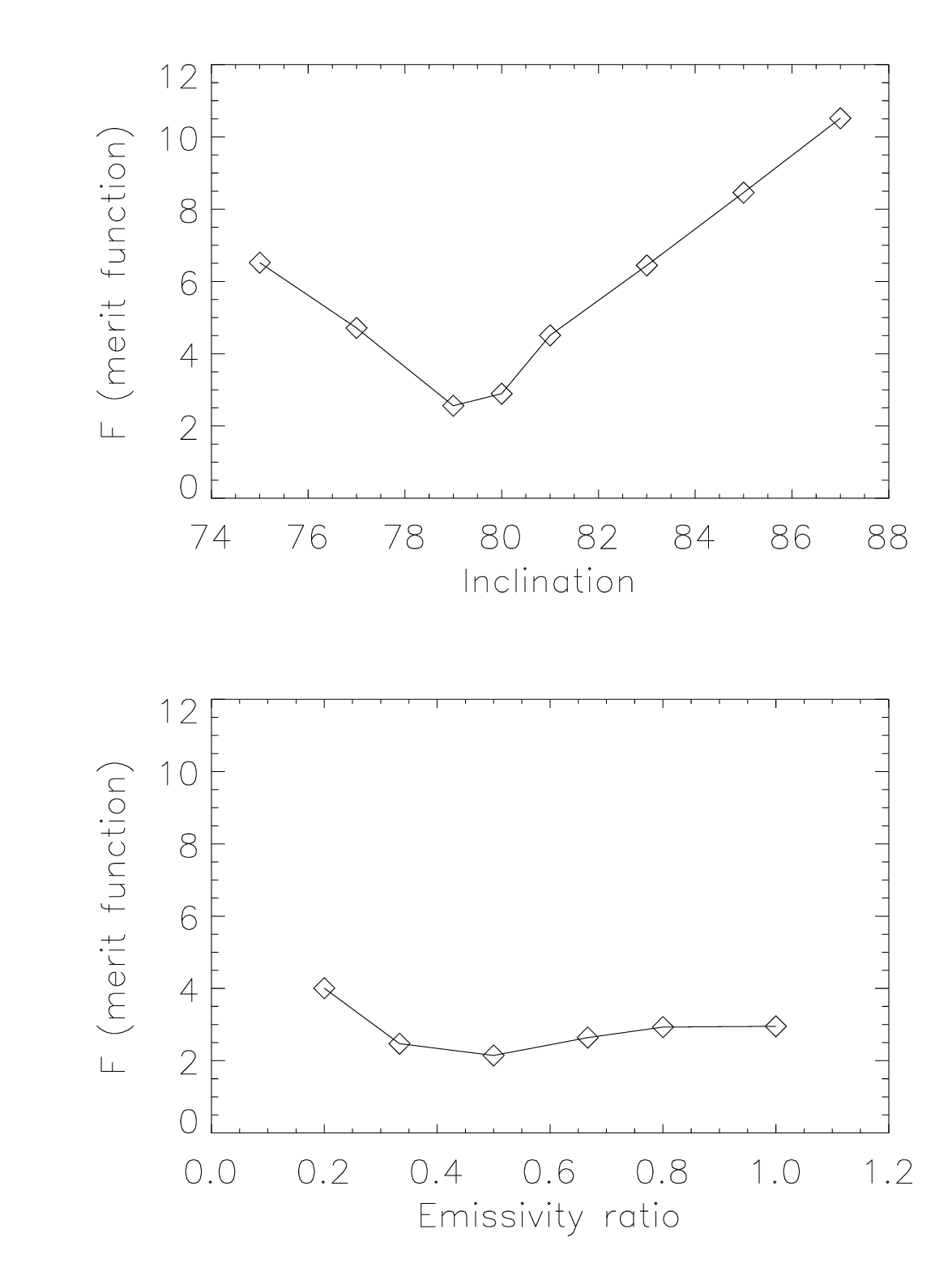,height=14.0cm,width=9.0cm,angle=0}}
\caption{The minimum values found for the merit function at given
fixed inclinations (top) and fixed emissivity ratios (bottom). This
was obtained using test set \#3, which has a known inclination of 80
degrees and individual fly emissivity ratio of 0.5.}
\end{figure}

\subsection{Fixed Parameters}

In addition to the location of the fireflies, there are fixed
parameters in our modelling which will, in principal, effect our results. 
We discuss these in more detail.

\subsubsection{Effects of inclination}

The system inclination and mass ratio $q=M_2/M_1$ are not, in general, 
precisely known for all eclipsing systems. However, since we know the orbital
period and the eclipse duration, the mass ratio is a function of
inclination only. So it is sufficient to study how the results depend on
using incorrect fixed inclinations for the modelling. 

As earlier, we have used our synthetic stream \#3 as a testbed for
our simulations. This set has an inclination of 80$^{\circ}$. We proceed
to fit the corresponding eclipse profile with 8 different inclination
values (fixed during each fit) ranging from 75$^{\circ}$ to 87$^{\circ}$. The
results are plotted in Figure 8 (top panel). Clearly, by fitting eclipse
profiles using different inclinations, the change in merit function
allows the inclination to be constrained by the fitting procedure. 

We could leave the inclination as a free parameter 
in our fits. However, its nature is different from the other fitted parameters
(the three degrees of freedom per firefly), and this would require 
a somewhat different fitting
procedure. As it is known for some systems, we have opted to 
leave inclination as a fixed input parameter. 

\subsubsection{Effects of angular emissivity function}

Secondly, we have studied the effect of the choice of emission 
law (equation 1). As in the case of inclination above, we have fitted 
set \#3, which was created with $F_{0} = 3.0, A = 1.0$ (white dwarf side
twice as bright as the other side, emissivity ratio = 0.5), 
with various values corresponding to the emissivity ratios (ER) of 0.2 to
1.0.  Surprisingly, there was no significant difference in the
quality of the fits, with the exception of ER = 0.2,
which gave a somewhat poorer fit. A plot of merit function versus 
fixed emissivity ratio is shown in Figure 8 (lower panel).  

The case of an optically thin
stream (and no X-ray heating, ER=1.0) produced a good fit, and the
resulting stream structure is almost unchanged from the ($F_{0} = 3.0, A =
1.0$, ER=0.5) case.  The implication is that unless the
emissivity function is known, a range of emissivity ratio must be explored to
separate out the apparent as opposed to the angle-independent brightness of the
stream. On the other hand the choice of 'wrong' emission law mainly
leads to poorer fits, but the resulting stream structure is still very
much like the one obtained using the correct emission law. 
This may be a consequence of a relatively narrow orbital phase range
($\sim$ 40$^{o}$) used in our fits (eg. the emission law would play a 
more significant role in a case where one would try to fit whole light
curves of these systems) . 

\subsection{Single pole versus two pole emission}

It is known, that some polars only accrete on to a single
pole, whilst other accrete on to both poles (Cropper 1990). Some systems 
can even switch between single and two pole accretion modes. Thus it
is worth while considering the effect of this on our modelling. 

As our regularisation term favours solutions where the accretion
stream consists of a single line or curve like structure, it is clear
that such regularisation cannot be correct for two pole emission,
where the stream divides. In order to model two pole accretion, we have
experimented with regularisation lines that, instead of starting
from L1 and ending on white dwarf, will both start AND end on the
white dwarf. This
would mimic a closed magnetic loop from one pole to another and,
in theory, would enable us to find the most probable magnetic loop 
like emitting stream. However, tests with such synthetic streams
did not prove very successful, and thus we have to limit our
modelling to single pole accretion case for the time being.
Further work on regularisation approach is needed to study 
this in depth.

\section{Conclusions}

We have presented a novel method to map the 3-dimensional accretion
stream structure in eclipsing AM Herculis systems. Our approach is the 
first in which no {\it a priori} information is assumed about the trajectory of
the accretion stream. This is important, as it allows us to study the
possible changes in accretion stream trajectory and accretion region location 
at different epochs, in different accretion states and in different
systems. Furthermore, by determining the trajectory we will gain
insight into the location of the magnetospheric threading region, 
and especially its distance from the white dwarf. This information, 
together with some assumptions of the stream density obtainable from 
the observations of X-ray dips (Watson \ea, 1989), could provide us 
with more evidence on the accretion rate.
  
Our tests show that the inversion method is sufficiently robust
to be applied to real data. This will be the subject of a forthcoming paper.

\section{Acknowledgements}

PJH is an Academy of Finland Research Fellow.

\newpage

\appendix

\section{Regularization using selforganizing maps}

Selforganizing maps (SOM) are a type of unsupervised neural networks
(Kohonen 1990). Their main use is data mining and cluster analysis of
multidimensional data. Due to their topology conserving nature, they
can also be used to measure similarities within data samples. In our 
approach we have used them to find a best fit curve through any given
swarm of fireflies, which is then used as a reference for computing the 
regularisation function for that particular swarm (i.e. we find a
smooth line through the swarm using SOM and compute the distances of 
individual flies to this line) .

SOMs consist of a grid of `neurons', which in the general case could be 
N-dimensional, but in our case it is 1-dimensional. Each of the
neurons has a set of weights associated with it. The number of weights
per neuron is exactly the same as the dimensionality of the input
data (in our case this means that the SOM 'net' consists of 20 neurons
(defining the smooth line) and each neuron has three weights, which
are simply its XYZ coordinates).
 
In a trained network, these weights will directly define the 
coordinates of the cluster centre associated with that particular
neuron (i.e. the location of the neuron in data space). To start with, 
these weights can have either random values
or they can span some predefined space otherwise. Here we chose to
initialize the weights by placing the neurons equidistantly along the
line from white dwarf to L1 point.

\subsection{Training SOM's}

The training of the network proceeds as follows. First a random data 
point ${\bf P}$ is chosen from the training sample of data points, in our case 
this would mean a selection of a random firefly from a swarm. Next, one
computes for each neuron ${\bf N}_{i}$, that has a set of $j$ weights
\begin{equation}
{\bf D}_{\rmn i} = \Big\|{\bf P} - {\bf N}_{\rmn i}\Big\|, \forall {\rmn i},
\end{equation}
i.e. the euclidean distance between the sample data point ${\bf P}$ (single
firefly) and the neuron ${\bf N}_{\rmn i}$. Then we define the winning neuron
${\bf N}_{\rmn w}$ as the one that has minimum ${\bf D}_{\rmn i}$ i.e. is the closest one to 
the input point ${\bf P}$ (chosen firefly). 

Next, the network is trained. This means simply, that for any
given swarm: We select a random fly, find the nearest neuron to it in 3D
space (A1), move that neuron (and it's neighbouring neurons as defined
below) towards the location of that fly (A2), and then proceed to the next
random fly from our swarm. This is iterated for about 5000 times (every
fly gets selected on the average about 25 times) and as a result the
neurons define a smooth curve through the swarm. 

Formally this happens as follows:

\begin{equation}
{\bf N}_{\rmn i}(t+1)={\bf N}_{\rmn i}(t)+h(i,w)({\bf P}-{\bf N}_{\rmn i}(t)), \forall {\rmn i},
\end{equation}
where ${\bf N}_{\rmn i}(t+1)$ is the updated neuron weight vector and 
$h(i,w)$ is a neigbourhood kernel, which depends on the distance of neuron
${\bf N}_{\rmn i}$ from the winning neuron ${\bf N}_{\rmn w}$ in the following manner:
\begin{equation}
h(i,w)=\alpha(t)e^{-\|{\bf r}_{\rmn i}-{\bf r}_{\rmn w}\|^2 \over 2\sigma(t)^2},
\forall {\rmn i}.
\end{equation}
Here, $\alpha(t)$ is a time dependant learning rate,
$\|{\bf r}_{\rmn i}-{\bf r}_{\rmn w}\|^2$ is the squared distance between the
winning Neuron ${\bf N}_{\rmn w}$
and the current neuron to be trained ${\bf N}_{\rmn i}$. Note that this distance
is the distance between the two neurons on the grid, and not the
distance between their weights. $\sigma(t)$ is the time dependent width
of the kernel. Other forms than the Gaussian dependance for $h(i,w)$
are also possible. The effect of the neighbourhood kernel is that
whilst the weights of the winning neuron are moved towards the input
data also the weights of its neighbouring neurons (on the grid) are
moved, but with smaller amount. This ensures the topology conserving
mapping, typical to the SOM.    

This completes one training step. Next one selects a new random input
and iterates again. The learning rate $\alpha(t)$ and the kernel width
$\sigma(t)$ are both constantly lowered during the training.

\subsection{Application to regularisation}

As a result of training, the SOM will find a predefined number of clusters
in the input data. For our purposes we are not really interested in
the clustering, but use the SOM to find a predefined ordered set of
nodes that define a smooth curve through our swarm of fireflies. This
is achieved by training a 1-dimensional SOM with fireflies as input
points. In addition, about 20-30 \% of the input points
are not fireflies but contain either the coordinates of the white dwarf
or the L1-point. We also give a constraint that the `white dwarf points' 
should 
be mapped to the other end of the map and the `L1-points' to the
other. This then produces a curve that stretches from
near the L1-point, through the swarm and terminates near the white dwarf. 

The regularisation itself means that we prefer solutions that are
well represented with the curve computed as above. Formally, we
use a SOM with 20 neurons, weights of which originally span the 
distance from L1 point to the white dwarf equally spaced. After
training the net with about 5000 inputs randomly selected as explained
earlier, we have the default path defined for the given swarm. We now
compute the regularisation term $S_{\rmn reg}$ as a sum of squared distances from 
this default path for all our 200 fireflies:
\begin{equation}
S_{\rmn reg}=\sum_{\rmn k=1}^{200}{({\bf P}_{\rmn k}-{\bf N}_{\rmn w,k})^2}.
\end{equation}
Here ${\bf P}_{\rmn k}$ is a vector in 3D space representing the location of
a single firefly and ${\bf N}_{\rmn w,k}$ is the location of the
nearest (to the ${\bf P}_{\rmn k}$ ) neuron .
This type of regularisation prefers solutions that are `stream-like'
with fixed ends, but leave the stream path totally free.        

\section{Genetic Optimisation}

The Genetic Algorithm (GA) used here for optimisation is similar
to those used in Hakala (1995), Harrop-Allin, Hakala \& Cropper (1999)
and Potter, Hakala \& Cropper (1998). There are, however, some
alterations specific to this procedure. In our current approach 
the ''genes'' that define the solution swarm are the locations of the
individual fireflies.  

In general, the algorithm proceeds as follows. 
First we create a a population of $\sim$ 500 random solutions 
{\it i.e.} 500 random swarms inside the Roche lobe of the
primary. Secondly, we evaluate these solutions.
Next, we select the parent solutions to produce offspring. 
This is done by first selecting 5 random solutions, the fittest of which
is chosen as 1 parent. This is repeated to choose the second parent.
Two offspring solutions are produced by `mixing' the parent swarms. In
practice this is done by picking fireflies at random from the two parent
swarms. Offspring solutions are mutated in three different ways. Firstly,
at a small probability of $\sim$ 0.002 any firefly within the swarm can be
removed and replaced by a new random firefly. With slightly higher
probability, a firefly can be mutated by changing its location by a small
gaussian random number. Finally a small fraction ($\sim 5 \%$) of the 
fireflies in the resulting swarm are mutated by moving them slightly 
towards the regularisation curve for that particular swarm (computed
using the SOM method as explained above). This speeds up the process when it 
otherwise gets very inefficient towards the latter stages of fitting.  

Having created offspring solutions we then evaluate them and if they
are better than their parent solutions they will replace the parents in the
population.
This procedure is repeated until some predefined convergence criterium 
is achieved or some fixed number of generations has passed.    

\end{document}